\documentclass[titlepage,12pt]{utarticle}

\usepackage{amsmath}
\usepackage{mathrsfs}
\long\def\omit#1{}

\newcommand\HH{\mathcal{H}}

\newcommand\MM{\mathcal{M}}

\newcommand\OO{\mathcal{O}}

\newcommand\dM{\partial \MM}

\newcommand\eps{\epsilon}

\newcommand\nts{\negthickspace}

\newcommand\tg{\tilde{g}}

\newcommand\rh{r_{_{H}}}
\newcommand\rw{r_{_{W}}}

\DeclareMathAlphabet{\mathpzc}{OT1}{pzc}{m}{it}

\begin{document}

\preprint{MCTP--04-65\\ {\tt hep-th/0411121}\\ }

\title{Boundary Counterterms and the Thermodynamics of 2-D Black Holes}

\author{Joshua L. Davis\footnote{\texttt{joshuald@umich.edu}} ~and Robert  
McNees\footnote{\texttt{ramcnees@umich.edu}}}

\oneaddress{Michigan Center for Theoretical Physics\\
  University of Michigan\\
        Ann Arbor, MI-48109, USA  }

\date{}

\Abstract{We utilize a novel method to study the thermodynamics of two dimensional type 0A black holes with constant RR flux. Our approach is based on the Hamilton-Jacobi method of deriving boundary counterterms. We demonstrate this approach by recovering the standard results for a well understood example, Witten's black hole. Between this example and the 0A black hole we find universal expressions for the entropy and black hole mass, as well as the infra-red divergence of the partition function. As a non-trivial check of our results we verify the first law of thermodynamics for these systems. Our results for the mass disagree with the predictions of a proposed matrix model dual of the 0A black hole.}

\maketitle

\section{Introduction}
\label{sec:introduction}

In the last few years, there has been a renewed interest in two-dimensional physics, spurred by advances in our understanding of the $c=1$ matrix model (for reviews, see \cite{Klebanov:1991qa,Polchinski:1994mb,Ginsparg:1993is}), particularly regarding the role of D-branes \cite{McGreevy:2003kb}. This recent work has revisited several two-dimensional examples of the early 1990's, which are largely related to the bosonic string, and also explored new physical systems associated with (worldsheet) supersymmetric string theories. These systems have contributed significantly to the discussion by providing new, non-perturbatively stable realizations of open/closed string duality. In addition these theories also have a richer low-energy spectrum than the bosonic string, including fermions and fluxes\cite{Douglas:2003up,Takayanagi:2003sm,Takayanagi:2004ge}.

Most effort has been devoted to two types of time-independent backgrounds: ``flat space'' and black holes. The ``flat space'' backgrounds are well-described by worldsheet methods. They have a trivial target space metric and a spacelike linear dilaton. The linear dilaton indicates that there is an asymptotic region which is weakly coupled, and one which is strongly coupled. Typically the strong coupling region is classically forbidden due to a large potential. The observable processes of the theory involve scattering of quanta off of this potential. There has been a great deal of success in describing string theory in these backgrounds with matrix models\cite{Douglas:2003up,Takayanagi:2003sm}. 

On the other hand, the status of two-dimensional black holes has remained somewhat murky. There exists a solution to bosonic string theory, wherein the worldsheet dynamics are described by a $SL(2,{\mathbb{R}})/U(1)$ coset model\cite{Witten:1991yr}. This solution is known variably as the $D=2$ Euclidean black hole, the cigar geometry, or Witten's black hole. In the limit $k \to \infty$, where $k$ is the level of the worldsheet current algebra, the worldsheet theory is weakly coupled and one can perform a semi-classical analysis. The large level limit of the coset model is also a solution to the tree-level two-dimensional beta function equations at lowest order in $\alpha^\prime$. We will be concerned with Witten's black hole in the limit where the lowest order beta-function equations are valid. In addition, there are a handful of other two-dimensional black hole solutions which are not known to have a worldsheet description. In this paper we will focus on one such object, the 0A black hole with constant RR flux, a solution to the genus zero beta function equations of type 0A string theory to leading order in $\alpha^\prime$\cite{Berkovits:2001tg}.

The primary goal of this paper is a study of the thermodynamics of the 0A black hole solution. A number of attempts have been made in this direction \cite{Berkovits:2001tg,Gukov:2003yp,Davis:2004xb,Danielsson:2004xf}, with conflicting results. The difficulty lies in the lack of a natural scheme for removing infra-red divergences from the Euclidean action for the solution. A common approach, background subtraction, was used to study the thermodynamics of the Witten black hole in \cite{Gibbons:1992rh,McGuigan:1991qp,Nappi:1992as} and the 0A black hole in \cite{Davis:2004xb,Danielsson:2004xf}. 
We will use a different technique, based on the Hamilton-Jacobi method for determining boundary counterterms \cite{Balasubramanian:1999re,Emparan:1999pm,deBoer:1999xf,deBoer:2000cz,Martelli:2002sp}. This method is intrinsic to the spacetime; it does not require a reference spacetime like background subtraction.
The summary of the technique is as follows. The action is regulated by truncating the spacetime at some large coordinate distance from the black hole horizon. This effectively adds a `regulating boundary' to the spacetime. We postulate an additional term in the action, which is supported only on the regulating boundary. This additional term must be intrinsic to the regulating boundary so that the bulk equations of motion go unmodified. The form of this additional ``counterterm'' Lagrangian is determined using three criteria. It should comprise local terms intrinsic to the boundary, it should have the same symmetries as the original action, and it should solve the Hamilton-Jacobi equation obtained from the Hamiltonian constraint for the action. The resulting counterterm action characterizes the infra-red divergences of the on-shell action, which can then be removed to produce a renormalized action. Once we have the renormalized action, standard techniques allow us to calculate the full thermodynamics of the black hole. We verify this approach by confirming the results of \cite{Gibbons:1992rh}, as well as previous calculations of the 0A black hole entropy \cite{Gukov:2003yp,Davis:2004xb, Danielsson:2004xf}. However, the mass we obtain for the 0A black hole is a new result. We confirm our result by showing that it satisfies the first law of black hole thermodynamics. To the best of our knowledge, this check has not been performed in the other analyses of these spacetimes.

As stated previously, the study of two-dimesional black holes in string theory has not been as fruitful as that of the two-dimensional flat space models. With a few exceptions (most notably the coset model description of Witten's black hole), one does not have worldsheet descriptions of these objects. Although there have been a number of attempts at providing a matrix model for a two-dimensional black hole \cite{Kazakov:2000pm,Jevicki:1993zg}, none of the proposals have enjoyed the success of the $c=1$ model or its supersymmetric generalizations. In fact, doubts have been expressed as to whether black holes can be formed in two-dimensional string theory \cite{Martinec:2004qt,Friess:2004tq}, or even exist as eternal solutions. These discussions are outside the scope of this paper, and we will not comment on them here. 

The structure of this paper is as follows. In section 2, we will summarize the solutions that we will be studying throughout the paper. Then, in section 3, we calculate the on-shell action for each solution. In section 4, we will discuss how the variational principle used to calculate the on-shell action determines the thermodynamic ensemble being studied. In section 5 we outline our technique for deriving boundary counterterms and find a universal form for the counterterms for both of our examples. Section 6 contains the actual thermodynamic calculations and verification of the first law. Finally we close with some discussion in section 7.

\section{Black Holes in Two Dimensions}
\label{sec:2DBH}

In this section we will summarize the derivation and properties of two black hole solutions. First we will construct the $k \rightarrow \infty$ black hole solution of \cite{Witten:1991yr} in two dimensional dilaton gravity. We will then derive a similar solution in two dimensional type 0A string theory with constant R-R flux. The explicit construction of these solutions is intended, in part, to clarify the relation between our conventions and others in the literature.

\subsection{Witten's Black Hole}
\label{subsec:WittenBH}

The lowest order $\beta$-functions for the two dimensional bosonic string can be derived from an effective space-time action of the form:
\begin{eqnarray} \label{bosonic_action}
  I & = & -\frac{1}{2 \kappa^2}\, \int_{\MM} \nts d^2x\, \sqrt{g} \,e^{-2\phi}\,\left(c + R + 4 (\nabla \phi)^2 
  	- (\nabla T)^2 -V(T) \right) \\ \nonumber
    &   & -\frac{1}{\kappa^2} \int_{\dM} \nts dx \,\sqrt{\tg} \,e^{-2\phi} \,K
\end{eqnarray}
In this action $c=16/\alpha'$ is related to the ``excess" central charge, and $T$ is the tachyon. Note that we have included the usual Gibbons-Hawking boundary term \cite{Gibbons:1976ue} in the action, which is required for the action to have a well defined variational principle.
 
We are interested in solutions of this theory where the tachyon vanishes, in which case the equations of motion are:
\begin{eqnarray} \label{bosonic_dilaton_eom}
	c + R + 4 \nabla^{2} \phi -4 (\nabla \phi)^2 & = & 0 \\ \label{bosonic_metric_eom}
	R_{\mu\nu} + 2 \nabla_{\mu} \nabla_{\nu} \phi & = & 0 
\end{eqnarray}
The linear dilaton, black hole solution \cite{Witten:1991yr} of these equations can be written:
\begin{eqnarray} \label{witten_bh_soln}
	ds^2 & = & -l(r)\, dt^2 + \frac{1}{l(r)}\,dr^2 \\
	\phi & = & -\frac{1}{2} \,\sqrt{c}\, \left( r-\rh \right) + \phi_{H}
\end{eqnarray}
In this solution $\rh$ denotes the location of the horizon,  $\phi_{H}$ is the value of the dilaton at the horizon, and the function $l(r)$ is:
\begin{eqnarray}\label{bosonic_l}
l(r) & = & 1 - e^{-\sqrt{c} \,( r-\rh) }
\end{eqnarray}
An asymptotic observer measures a temperature for the black hole given by:
\begin{equation}\label{bosonic_T_c_reln}
    T \,\, = \,\, \frac{1}{4\pi}\, l'(\rh) \,\, = \,\, \frac{\sqrt{c}}{\,4\pi}
\end{equation}
The thermodynamics of this black hole is well-understood \cite{Gibbons:1992rh,McGuigan:1991qp, Nappi:1992as}. Therefore, it provides a nice example for illustrating our approach before addressing the 0A black hole, which we now review.

\subsection{Black Hole Solutions Of Type 0A Supergravity} \label{sec:0ABH}
  The effective low energy action of type 0A string theory in two dimensions is given by \cite{Douglas:2003up}:
\begin{eqnarray}\label{0Aaction}
  I & = & -\int_{\MM} d^2x \sqrt{g} \left[ \frac{1}{2 \kappa^2}\,e^{-2\phi}\left(c+R+4(\nabla \phi)^2 - a (\nabla T)^2 
  +\frac{2a}{\alpha'}\,T^2 +\ldots\right)\right. \\ \nonumber
    &   & \hspace{20pt} \left. -\frac{2\pi\alpha'}{4}\,\left(e^{-2T}|F^{(-)}|^2+e^{2T}|F^{(+)}|^2\right)
    		+\ldots\right] \\ \nonumber
    &   & -\frac{1}{\kappa^2} \int_{\dM}dx \sqrt{\tg} \,e^{-2\phi} K
\end{eqnarray}
This action describes the metric, dilaton, and tachyon, as well as two Ramond-Ramond gauge fields. In the two-dimensional theory $c$ is given by $8/\alpha'$. We are once again interested in solutions with vanishing tachyon. Consistently setting the tachyon to zero requires that we cancel terms linear in $T$ in the action \eqref{0Aaction}. This means that the RR field strengths must be equal. In this case the action reduces to:
\begin{eqnarray}
  I & = & -\int_{\MM} d^2x \sqrt{g} \left[ \frac{1}{2 \kappa^2}\,e^{-2\phi}\left(\frac{8}{\alpha'}+R
  		+4(\nabla \phi)^2\right)-\frac{1}{2} \, (2\pi \alpha') F^{\mu\nu} F_{\mu\nu}\right] \\ \nonumber
    &   & -\frac{1}{\kappa^2} \int_{\dM}dx \sqrt{\tg} \,e^{-2\phi} K
\end{eqnarray}
In the following we will set $2\kappa^2 = 1$. Note that, because the two RR field strengths must have the same flux, we have written our action in terms of a single field strength
with a non-canonical coefficient. Once we begin to analyze the black hole thermodynamics it will be important to remember that this field strength actually represents two canonically normalized field strengths with equal flux. 

It is straightforward to vary the action \eqref{0Aaction} and obtain the following equations of motion for the gauge field and the dilaton:
\begin{eqnarray}
 \nabla_{\mu}F^{\mu\nu} & = & 0
\end{eqnarray}
\begin{eqnarray}
 c + R + 4 \nabla^{2}\phi - 4 \left( \nabla \phi\right)^2 & = & 0
\end{eqnarray}
Varying the action with respect to the metric yields the Einstein Equations:
\begin{eqnarray}
  G_{\mu\nu} & = & T_{\mu\nu}
\end{eqnarray}
In two dimensions the Einstein tensor vanishes identically due to the identity $R_{\mu\nu} = \frac{1}{2}\,g_{\mu\nu} R$, so the Einstein equations become $T_{\mu\nu} = 0$. The stress tensor is given by:
\begin{eqnarray}
  T_{\mu\nu} & = & \frac{1}{2}\,g_{\mu\nu} \,e^{-2\phi} \left( c+4\nabla^{2}\phi-4\left(\nabla 
  \phi\right)^2\right)-2e^{-2\phi}\nabla_{\mu}\nabla_{\nu} \phi \\ \nonumber
  	     &   & + (2\pi\alpha')\,\left( F_{\mu}^{\,\lambda}F_{\nu\lambda} - 
	     \frac{1}{4}\,g_{\mu\nu}\,F^{\lambda\rho}F_{\lambda\rho}\right)
\end{eqnarray}
We now look for a solution of these equations that corresponds to a black hole in a linear dilaton background, with constant R-R flux. We choose the $r \rightarrow \infty$ limit to correspond to asymptotically flat spacetime, and require that the function $g_{tt}(r)$ vanishes at a horizon $r=\rh$. This leads us to the following solution:
\begin{eqnarray} \label{bhsoln}
  ds^{2} & = & -l(r) \,dt^2 + \frac{1}{l(r)} \,dr^2 \\ \nonumber
  F_{\mu\nu} & = & \frac{q}{2\pi\alpha'}\,\eps_{\mu\nu} \\ \nonumber
  \phi(r) & = & -\frac{1}{2}\sqrt{c} \,(r-\rh)+\phi_{H}
\end{eqnarray}
The function $l(r)$ appearing in the metric is given by:
\begin{eqnarray}\label{0A_l}
  l(r) & = & 1-e^{-\sqrt{c}\,(r-\rh)}-\frac{q^2}{2\pi\alpha'}\,\frac{1}{\sqrt{c}}\,(r-\rh) e^{-\sqrt{c}\,(r-\rh)+2\phi_{H}}
\end{eqnarray}
In the $q \rightarrow 0$ limit this solution takes the same form as the black hole of the previous section.

The horizon of the black hole solution \eqref{bhsoln} is located at $r=\rh$. The black hole temperature is given by:
\begin{eqnarray}
	T & = & \frac{1}{4\pi}\,l'(\rh)
\end{eqnarray}
In terms of $c$, $q$, and $\phi_{_H}$ this is:
\begin{eqnarray}\label{0A_temp_reln}
	4 \pi T & = & \sqrt{c} - \frac{q^2}{2\pi\alpha'}\,\frac{1}{\sqrt{c}}\,e^{2\phi_H}
\end{eqnarray}
The $T \rightarrow 0$ limit sets a maximum value for the electric charge
. In terms of $c$ and the value of the dilaton at the horizon this is:
\begin{eqnarray}
	q_{max}^{\,2} & = & 2\pi\alpha'\,c\,e^{-2 \phi_H}
\end{eqnarray}
Alternately, we can express this in terms of the string coupling at the horizon as $q_{max} = 16 \pi \, g_{H}^{\,-2}$.
This upper limit represents the extremal black hole, where both $l(r)$ and $l'(r)$ vanish at the horizon.

The field strenth is associated with a $U(1)$ gauge field $A_{\mu}$. We partially fix the gauge by setting $A_{r} = 0$. The remaining component $A_t$ is then determined by $F_{rt} = \partial_r A_{t}$, which gives:
\begin{eqnarray} \label{gauge_field}
   A_{t}(r) & = & \frac{q}{2\pi \alpha'}\,r + a 
\end{eqnarray}
where $a$ is an undetermined constant. There is a residual gauge freedom associated with gauge transformations $A_{\mu} \rightarrow A_{\mu} + \partial_{\mu} \Lambda(t)$ which preserves the original gauge choice $A_{r} = 0$. We will eliminate this remaining invariance later on and completely fix the gauge. 

Finally, we should mention that in several references \cite{Berkovits:2001tg,Gukov:2003yp,Davis:2004xb} the function $l(r)$ appearing in the metric is written (after suitable rescalings) in terms of a `mass parameter' $m$ as:
\begin{eqnarray}\label{alternate_0A_metric}
  l(r) & = & 1-\frac{4}{c}\,e^{-\sqrt{c}\,r}\,\left( m + \frac{\sqrt{c}}{4}\,\frac{q^2}{2\pi\alpha'}\,r\right)
\end{eqnarray}
For the sake of comparison, this is related to our expression \eqref{0A_l} by:
\begin{eqnarray}\label{extremality_parameter}
	4\frac{m}{\sqrt{c}} & = & \sqrt{c}\,e^{-2\phi_H}
		-\frac{q^2}{2\pi\alpha'}\,\rh
\end{eqnarray}

\section{The Partition Function and On-Shell Actions}\label{sec:On_Shell_Actions}
In this section we briefly review the general procedure for obtaining the thermodynamic partition function in terms of the on-shell gravitational action. We then calculate the on-shell action for each of the solutions discussed in section \ref{sec:2DBH}. The variational procedure used to calculate the on-shell action will lead us to a discussion of the relevant thermodynamic ensembles in the following section. 

\subsection{The Partition Function}
In the usual approach to black hole thermodynamics the partition function is given by a path integral weighted by the exponential of the Euclidean action. We assume that the path integral is dominated by solutions close to the classical field configurations, so that the partition function can be approximated as
\begin{eqnarray}\label{partition_function}
    Z & = & \exp \left( - I_{E}\right)
\end{eqnarray}
The Euclidean action appearing in \eqref{partition_function} is the on-shell action, evaluated for Euclideanized field configurations solving the bulk equations of motion. In practice we will evaluate the partition function and the relevant thermodynamic potentials in terms of the Euclidean action, but present them in terms of quantities which have been Wick rotated back to Lorentzian signature. This will not lead to any confusion, as long as we remember to impose the appropriate periodicity and regularity conditions on the fields in the action. 

In the following two sub-sections we will evaluate the on-shell action for the black hole solutions presented in section \ref{sec:2DBH}. In higher dimensions it is well known that such gravitational actions contain infra-red divergences which must be addressed in order to obtain a sensible thermodynamics \cite{Gibbons:1976ue}. The issue is more subtle in two dimensions, and will be discussed in detail in section \ref{sec:Divergences}. For now, we avoid this issue simply by regulating calculations of the action with a `wall' placed at $r=\rw$. 
The boundary of the regulated spacetime is 
the one-dimensional hypersurface defined by
$r=\rw$ 
. Both solutions we consider have metrics with the same form:
\begin{eqnarray}
   ds^2 & = & -l(r) \,dt^2 + \frac{1}{l(r)}\,dr^2
\end{eqnarray}
so the unit normal that defines the boundary at $r=\rw$ is given in both cases by:
\begin{eqnarray}
   n_{\mu} \,\,=\,\, \delta_{\mu r} \frac{1}{\sqrt{l(r)}} & \,\, & n^{\mu} \,\,=\,\, \delta_{\mu r} \sqrt{l(r)} 
\end{eqnarray}
The trace of the extrinsic curvature, which appears in the action, is given by the covariant divergence of this normal:
\begin{equation}
   K \,\, = \, \, \nabla_{\mu} n^{\mu} \, \, = \, \, \frac{l'(r)}{2 \sqrt{l(r)}}
\end{equation}
The induced metric on the boundary consists of the single component $\tg_{tt} = g_{tt}$, so the metric factor in the measure on the boundary is simply $\sqrt{\tg} = \sqrt{-\tg_{tt}(\rw)}$.

\subsection{The Regulated Action for Witten's Black Hole}

When the tachyon is set to zero the action \eqref{bosonic_action} reduces to:
\begin{eqnarray}
  I & = & -\, \int_{\MM} \nts d^2x\, \sqrt{g} \,e^{-2\phi}\,\left(c + R + 4 (\nabla \phi)^2  \right) - 2 
  \int_{\dM} \nts dx \,\sqrt{\tg} \,e^{-2\phi} \,K
\end{eqnarray}
Using the dilaton equation of motion \eqref{bosonic_dilaton_eom} we can re-write the bulk integrand, which gives:
\begin{eqnarray}
  I & = & -\, \int_{\MM} \nts d^2x\, \sqrt{g} \,e^{-2\phi}\,\left(8(\nabla \phi)^2  -4 \nabla^{2} \phi \right) - 2 
  \int_{\dM} \nts dx \,\sqrt{\tg} \,e^{-2\phi} \,K
\end{eqnarray}
Integration by parts cancels the bulk terms, and leaves a boundary term:
\begin{eqnarray}
  I & = &  \int_{\dM} \nts dx \,\sqrt{\tg} \,e^{-2\phi} \,\left( 4 \,n_{\mu} \nabla^{\mu} \phi - 2\,K \right)
\end{eqnarray}
Evaluating the integrand for the linear dilaton black hole gives:
\begin{eqnarray}\label{onshell_integrand}
 \sqrt{\tg}\,e^{-2 \phi} \, \left( 4 \,n_{\mu} \nabla^{\mu} \phi - 2\,K \right) & = & - \, e^{-2\phi} 
 	\, \left(2 \sqrt{c} \, l(r) + l'(r)\right)
\end{eqnarray}
Using the expression \eqref{bosonic_l} for the function $l(r)$ appearing in the metric, the derivative in the last term is simply:
\begin{eqnarray}
	l'(r) & = & \sqrt{c}\,\left(1-l(r) \right) 
\end{eqnarray}
We are interested in the Euclidean action, in which case the imaginary time $\tau = i\,t$ has periodicity $\beta = T^{-1}$. Evaluating these boundary terms at the wall $r=\rw$ gives an action:
\begin{eqnarray} \label{WittenBH_Action}
   I_{E} & = & -\beta \, \sqrt{c} \, e^{-2 \phi(\rw)} \, \left( 1 + l(\rw) \right)
\end{eqnarray}

\subsection{The Regulated Action for the 0A Black Hole} 
The Euclidean action for the 0A black hole of section \ref{sec:0ABH} can be evaluated in the same manner as in the previous section. It differs only in the definition \eqref{0A_l} of the function $l(r)$ appearing in the metric, and the presence of the RR field strength, which does not couple to the dilaton. Thus, the form of the on-shell action is simply the integral of \eqref{onshell_integrand} over the boundary, plus the RR term:
\begin{eqnarray}\label{RR_action}
   I_{RR} & = & \frac{1}{2} \, (2 \pi \alpha')\,\int_{\MM} \nts d^{2}x \sqrt{g} \,  F^{\mu\nu} F_{\mu\nu} 
\end{eqnarray}
We begin by rewriting the integrand and integrating by parts:
\begin{equation}
   F^{\mu\nu} F_{\mu\nu} \,\, = \,\, 2\,F^{\mu\nu}\,\nabla_{\mu} A_{\nu} \,\, = \,\, 2\,\nabla_{\mu} \left(F^{\mu\nu} \, 
   	A_{\nu}\right) - 2 A_{\nu} \nabla_{\mu} F^{\mu\nu} 
\end{equation}
Thus, the contribution to the action \eqref{RR_action} becomes:
\begin{eqnarray}
 I_{RR} & = &  (2\pi\alpha')\,\int_{\dM} \nts dx \sqrt{\tg} \,  n_{\mu}\, A_{\nu} \, F^{\mu\nu}
 	- (2\pi\alpha') \, \int_{\MM} \nts d^{2}x \sqrt{g} \, A_{\nu} \, \nabla_{\mu} F^{\mu\nu} 
\end{eqnarray}
The bulk term vanishes by the equations of motion, \emph{except} for at $r=\rh$ where the gauge field is not regular. However, the Euclidean path integral should only involve fields with appropriate periodicities and regularity at the origin. In our choice of gauge the gauge field \eqref{gauge_field} is of the form:
\begin{eqnarray}
	A_{\mu} & \sim & \partial_{\mu} \tau 
\end{eqnarray}
where $\tau$, the imaginary time, is an angular coordinate about the `axis' $r=\rh$. To obtain a regular gauge field at the origin of this coordinate system, where $\tau$ is undefined, we use the remaining gauge freedom to set the constant $a$ appearing in \eqref{gauge_field} so that the electric potential vanishes at the horizon:
\begin{eqnarray} \label{regular_gauge_field}
  A_{t}(r) & = & \frac{q}{2\pi\alpha'}\,\left( r-\rh \right)
\end{eqnarray}
We can now evaluate the RR contribution to the on-shell action. The bulk term vanishes everywhere on-shell, leaving only a boundary term:
\begin{eqnarray}
 I_{RR} & = &  -\beta \, \frac{q^2}{2\pi \alpha'}\,\left( \rw-\rh \right)
\end{eqnarray}
Adding this contribution to  \eqref{onshell_integrand} yields:
\begin{eqnarray}
    I_{E} & = & -\beta \, e^{-2 \phi(\rw)} \, \left(2 \sqrt{c} \, l(\rw) + l'(\rw) \right) - \beta \, \frac{q^2}{2\pi 
    \alpha'}\,\left( \rw-\rh \right)
\end{eqnarray}
Rewriting the derivative of $l(r)$ using \eqref{0A_l} gives the on-shell Euclidean action for the $0A$ black hole:
\begin{eqnarray}\label{OA_reg_action}
    I_{E} & = & -\beta \, \sqrt{c} \,e^{-2 \phi(\rw)} \, \left( 1 +  l(\rw) \right) + \beta \, \frac{q^2}{2\pi \alpha' \sqrt{c}} - 
    \beta \, q \, \Phi
\end{eqnarray}
In the last term we define $\Phi$, the electrostatic potential between the wall and the horizon, as:
\begin{eqnarray}
	\Phi & \equiv & A_{t}(\rw) - A_{t}(\rh)
\end{eqnarray}

\subsection{ The Wall and Divergences}
 
Now that we have obtained the regulated Euclidean actions for the two black hole solutions of section \ref{sec:2DBH} we can clearly see the effect of removing the `wall' located at $r=\rw$. The functions $l(r)$ appearing in both actions asymptote to 1 as $\rw \rightarrow \infty$, which was simply our requirement that far away from the black hole we recover the usual Minkowski metric:
\begin{eqnarray}
  \lim_{\rw \rightarrow \infty} g_{\mu\nu} & = & \eta_{\mu\nu} 
\end{eqnarray}
In the same limit the string coupling goes to zero, which means that the exponential dilaton factor $e^{-2\phi(\rw)}$ diverges. In section \ref{sec:Divergences} we will discuss these divergences and our approach to removing them, which will lead to a partition function with sensible thermodynamics. But first we turn to the nature of the partition function, and the relation between the variational principle used to obtain the on-shell actions and the thermodynamic ensembles these actions represent.  

\section{Black Hole Thermodynamics and The Question of Ensembles}
\label{sec:Ensembles}

In this section we show how the variational principle used to extremize the gravitational action specifies the variables that the partition function depends on, and therefore determines the particular thermodynamic ensemble we are studying. We introduce the appropriate thermodynamic variables for each of the theories discussed in section \ref{sec:2DBH} and the formulae for determining relevant thermodynamic quantities.

\subsection{On-Shell Actions as Boundary Functionals}

The approximation \eqref{partition_function} for the partition function comes from evaluating the contributions to the path integral from field configurations `close' to the solutions of the classical equations of motion. In other words, if  $\phi_{cl}$ is a solution of the classical equation of motion associated with an action $I[\phi]$, then we assume that the path integral is dominated by fields $\phi = \phi_{cl} + \delta \phi$ that represent small deviations from the classical solution. The action for such fields can be expanded as:
\begin{eqnarray}\label{action_expansion}
  I[\phi_{cl}+\delta \phi] & = & I[\phi_{cl}] + \left. \frac{\delta I[\phi]}{\delta \phi} \right|_{\phi_{cl}}  \delta \phi
     +\frac{1}{2} \, \left. \frac{\delta^2 I[\phi]}{\delta^2 \phi} \right|_{\phi_{cl}} \delta \phi^{\,2} + \ldots
\end{eqnarray}
Since $\phi_{cl}$ solves the equations of motion, the linear term vanishes up to boundary terms coming from integration by parts. 
For these terms to vanish we fix the value of $\phi$ at the boundary by considering variations $\delta \phi$ which vanish there. Neglecting higher order terms in the expansion \eqref{action_expansion}, the on-shell action is approximately $I[\phi_{cl}]$.

Because \eqref{partition_function} is interpreted as the thermodynamic partition function for the black hole spacetime, the boundary conditions placed on the bulk fields determine the thermodynamic ensemble \cite{Brown:1989fa}. This is because the on-shell action, and therefore the partition function, is a function of the bulk fields evaluated at the boundary \footnote{ In general, the on-shell action is a \emph{functional} of the bulk fields evaluated at the boundary. The two dimensional examples we are considering are somewhat simpler, because the bulk fields are independent of the boundary coordinate $t$. Therefore, our on-shell actions are simply functions of the values that the bulk fields take on the regulating surface $r=\rw$.}. Depending on the theory being studied, the boundary value of a bulk field may be related either to a conserved charge, or to a chemical potential associated with a conserved charge. An example of the former is the dilaton, which is directly related to the dilaton charge, as described in the next subsection. An example of the latter is the time component of the gauge field, which is related to the electrostatic potential due to a conserved electric charge.

If fixing the bulk fields at the boundary corresponds to fixing conserved charges, then \eqref{partition_function} represents the partition function in the canonical ensemble. On the other hand, if fixing bulk fields at the boundary corresponds to holding chemical potentials fixed, then the partition function is calculated in the grand canonical ensemble. It is also possible, as we will see in the case of the 0A black hole, that the partition function is calculated in an ensemble which is canonical with respect to some variables, and grand canonical with respect to others.

\subsection{Witten's Black Hole}

For the solution in section \ref{subsec:WittenBH}, the value of the temperature is clear from the requirement that the imaginary time coordinate has the appropriate periodicity at the horizon. The other thermodynamic variable is the dilaton charge, which we denote by $D_W$. In two dimensions any function $f(\phi)$ of a scalar field leads to a conserved current:
\begin{eqnarray}
   j^{\mu} & = & \epsilon^{\mu\nu} \, \nabla_{\nu} f(\phi) 
\end{eqnarray}
Hence, there are an infinite number of conserved charges to choose from. We follow the usual convention \cite{Gibbons:1992rh} and choose the function $f(\phi) = \exp(-2\phi)$. This gives the square of the inverse string coupling at the wall as the dilaton charge $D_W$:
\begin{eqnarray} \label{dilaton_charge}
   D_W & = & e^{-2\phi(\rw)}
\end{eqnarray}
We write the dilaton charge with a subscript `$W$' to indicate its dependence on the location of the wall.

%

The variational principle used to obtain the equations of motion and the solution \eqref{witten_bh_soln} required that the metric and dilaton are fixed at the boundary of spacetime:
\begin{equation}
	\delta g_{\mu\nu} \, \, = \delta \phi \, \, = \, \, 0
\end{equation}
Given the definition \eqref{dilaton_charge}, this amounts to fixing the dilaton charge contained within the wall. Furthermore, in the path integral we fix the periodicity in the imaginary time at the horizon to be $\beta$, the inverse temperature. The conclusion is that we have computed the partition function in the canonical ensemble, producing a function of the temperature and the dilaton charge.  Since we have evaluated the boundary terms at $r=\rw$ they are properly expressed as functions of the blue-shifted temperature at the wall, $T_W$. This is given by the Tolman relationship \cite{Tolman:1934}:
\begin{eqnarray}\label{Tolman_factor}
	T_W & = & \frac{1}{\sqrt{-\tg_{tt}(\rw)}} \, T
\end{eqnarray}
The log of the partition function gives the Helmholtz free energy $F_W(T_W,D_W)$. The subscript reminds us that this quantity is the free energy for the finite volume system interior to the wall at $r=\rw$. We may then use the usual definitions (see, for example \cite{Reif:1965}) to obtain the entropy and dilaton potential (the chemical potential associated with the dilaton charge) of the system:
\begin{eqnarray}\label{canonical_defns}
	S & = & - \left. \frac{\partial F_W}{\partial T_W} \, \right\vert_{D_W} \\ \nonumber
	\psi_W & = &  \left. \frac{\partial F_W}{\partial D_W} \, \right\vert_{T_W}
\end{eqnarray}
Once we have the entropy, we can obtain the mass (energy) of the black hole spacetime by inverting the Legendre transform used to define the Helmholtz free energy:
\begin{eqnarray}
	M_W & = & F_W + T_W \, S
\end{eqnarray}

\subsection{The 0A Black Hole}
The reasoning in the previous section also applies to the case of the 0A black hole. The variational principle used to calculate the on-shell action leads to a partition function which is a function of the temperature at the wall, $T_W$, and the dilaton charge $D_W$. However, there is an additional conserved charge due to the RR gauge field(s). The partition function for the 0A black hole is not a function of the electric charge, which we denote by $Q$. Rather, it is a function of the electrostatic potential $\Phi_W$ between the wall and the horizon. This is because the potential is the time component of the gauge field $A_{\mu}$, which we have fixed at the boundary of spacetime when carrying out the variational principle that leads to the on-shell action. As a result we are working in the grand canonical ensemble with respect to the electric charge. Therefore, the log of the partition function corresponds to the Legendre transform of the Helmholtz free energy with respect to the electric charge, a quantity which we denote by $Y_W$:
\begin{eqnarray}
  Y_W(T_W,D_W,\Phi_W) & = & F_W (T_W,D_W,Q) - Q\,\Phi_W
\end{eqnarray}
The electrostatic potential $\Phi$ is related to the $t$ component of the gauge field, so the electrostatic potential that an asymptotic observer measures at the wall, $\Phi_W$, contains an appropriate Tolman factor:
\begin{eqnarray}\label{Phi_W_definition}
  \Phi_W & = & \frac{1}{\sqrt{-g_{tt}}}\,\Phi\\ \nonumber
         & = & \frac{T_W}{T} \, \frac{q}{2 \pi \alpha'} \, (\rw-\rh)
\end{eqnarray}
  
Although we are used to working in either the canonical or grand canonical ensemble, there is nothing wrong with considering a mix of the two. We are simply studying the thermodynamics of a system with fixed dilaton charge and electrostatic potential. It is straightforward to obtain the entropy, electric charge, and dilaton potential for the system:
\begin{eqnarray} \label{mixed_ensemble_defns}
	S & = & - \left. \frac{\partial Y_W}{\partial T_W} \, \right\vert_{D_W,\Phi_W} \\
	\psi_W & = &  \left. \frac{\partial Y_W}{\partial D_W} \, \right\vert_{T_W,\Phi_W} \\
	Q & = & - \left. \frac{\partial Y_W}{\partial \Phi_W} \, \right\vert_{T_W,D_W}
\end{eqnarray}
As in the previous section, once we have the entropy and the electric charge we can invert the Legendre transform used to define $Y_W$ and obtain the mass of the 0A black hole spacetime with constant RR flux:
\begin{eqnarray}
	M_W & = & Y_W + T_W \, S + Q \, \Phi_W 
\end{eqnarray}

We are now almost ready to calculate the thermodynamics of these solutions. The only thing that remains is to address the divergences that have been alluded to in previous sections. We discuss these in the next section.

\section{Taming Divergences}
\label{sec:Divergences}

In order to use the thermodynamic formulae of the previous section we must remove certain divergences from the on-shell actions, and hence the partition functions. In this section we review the nature of these divergences and summarize a method of systematically removing them from the on-shell action. In other words, a prescription for producing an appropriate renormalized action. This is illustrated using Witten's black hole, then applied to the 0A black hole.

\subsection{The Nature of the Problem}

There are many approaches to removing infra-red divergences from gravitational actions. The most common technique is `background subtraction' \cite{Gibbons:1976ue}, which was used in \cite{Gibbons:1992rh} to study the thermodynamics of Witten's black hole
\footnote{It was also used in \cite{Danielsson:2004xf} to study some aspects of the $0A$ black hole. Their calculation is performed in the canonical ensemble, which forces them to use a spacetime with charge $q$ in their background subtraction.}. Background subtraction, which uses a second, reference spacetime to identify divergences which should be subtracted from the action, has been remarkably successful. But in some cases the choice of reference spacetime is ambiguous. In those cases it would be nice to have a method that is intrinsic to the solution at hand, instead of one which requires another solution to compare to. 
 
In this paper we will use a generalization of the boundary counterterm method \cite{Balasubramanian:1999re,Emparan:1999pm}, often used to study black holes in asymptotically anti-de Sitter space \cite{Emparan:1999pm,Awad:1999xx,Awad:2000aj,Kraus:1999di,Buchel:2003re,Liu:2004it,Batrachenko:2004fd}. Our approach is based on the Hamilton-Jacobi method in which a counterterm action, made up of intrinsic boundary terms, is added to the on-shell action \cite{deBoer:1999xf,deBoer:2000cz,Martelli:2002sp}. The counterterm action cancels the appropriate infra-red divergences and gives a partition function with a well defined thermodynamics \footnote{Another approach, known as holographic renormalization \cite{Bianchi:2001kw}, exists. There are many reasons to believe that holographic renormalization and the Hamilton-Jacobi method are equivalent. We expect that applying it to the problem we study here would lead to the same results.}. 

Before describing this approach, it is important to note that the counterterm actions proposed in this section give results that are completely equivalent to what one obtains using the background subtraction method. The same arguments used by Gibbons and Perry in \cite{Gibbons:1992rh} can be used here, for both of the systems we consider. We employ the Hamilton-Jacobi approach for two reasons. First, as stated above, it appears to be a more general technique than background subtraction, and it is interesting to explore the range of problems to which it applies. Second, it leads to an elegant derivation of the first law of black hole thermodynamics as a consequence of Hamilton's Action principle. We will return to the second point in a subsequent publication \cite{Davis:2004xx}.

\subsection{Hamilton-Jacobi Counterterms}

The Hamilton-Jacobi approach has been reviewed in several publications; for a thorough discussion see \cite{Martelli:2002sp}. We will only give a brief outline of it here, using Witten's black hole as an example to illustrate the procedure. We will then verify that the renormalized action yields the correct thermodynamics \cite{Gibbons:1992rh}.
 
The regulated, on-shell actions calculated in section \ref{sec:On_Shell_Actions} contain terms which diverge in the $\rw \rightarrow \infty$ limit, as the regulating surface is taken to infinity. To remove these divergences we split the regulated, on-shell action into  a counterterm action $I_{ct}$ and a renormalized action $\Gamma$, as: 
\begin{eqnarray}
  I & = & I_{ct} + \Gamma
\end{eqnarray}
The renormalized action $\Gamma$ is the action we want to use to calculate the partition function. The counterterm action is a boundary integral whose integrand comprises a set of local terms on the regulating surface $r=\rw$. These terms are all intrinsic to the boundary, and therefore the variations of $I$ and $\Gamma$ lead to the same bulk equations of motion. 

The counterterm action is determined by requiring that it is a solution of the Hamilton-Jacobi equation associated with the Hamiltonian constraint, $\HH=0$, of the full action $I$. For example, for the action \eqref{bosonic_action} the Hamiltonian density  is given by:
\begin{eqnarray}\label{bosonic_Hamiltonian}
  \HH & = & \frac{1}{16} \, e^{2 \phi} \, \pi_{\phi}^{\,2}  - e^{2 \phi}
   	\, \left( \tg_{tt} \, \pi^{tt} + \frac{1}{4} \, \pi_{\phi} \right)^2 - e^{-2 \phi} \, c
\end{eqnarray}
The canonical momenta can be expressed as derivatives \footnote{In higher dimensional examples, the on-shell action is a functional of the bulk fields, evaluated at the boundary. In the two dimensional examples studied here, the bulk fields do not depend on the boundary coordinate $t$ and the on-shell action is therefore a function, as opposed to a functional, of the fields evaluated at the boundary.} of the on-shell action with respect to the fields at the boundary:
\begin{eqnarray}\label{bosonic_momenta}
   \pi_{\phi} & = & \frac{1}{\sqrt{\tg}} \, \frac{\partial \, I}{\partial \, \phi(\rw)} \\ \nonumber
   \pi^{tt}    & = & \frac{1}{\sqrt{\tg}} \, \frac{\partial \, I}{\partial \, \tg_{tt}(\rw)}
\end{eqnarray}
The Hamilton-Jacobi equation is simply the Hamiltonian constraint, expressed as a non-linear differential equation for the on-shell action. Requiring that the counterterm action also be a solution of the Hamilton-Jacobi equation gives:
\begin{eqnarray}\label{Hamilton_Jacobi_Equation}
\frac{1}{16} \, e^{2 \phi} \, \left(  \frac{\partial \, I_{ct}}{\partial \, \phi(\rw)}\right)^2  - e^{2 \phi}
   	\, \left( \tg_{tt} \, \left(  \frac{\partial \, I_{ct}}{\partial \, \tg_{tt}(\rw)}\right) + \frac{1}{4} \, 
	\left(  \frac{\partial \, I_{ct}}{\partial \, \phi(\rw)}\right) \right)^2 
	& = & -\tg_{tt}(\rw)\,e^{-2 \phi} \, c 
\end{eqnarray}
In order to solve this non-linear differential equation we make an ansatz for the counterterm action, based on the symmetries of the action.

First, we need a counterterm action that is invariant under diffeomorphisms of the single coordinate along the boundary. We enforce this requirement by writing the counterterm action as a boundary integral with the proper covariant volume element on the boundary:
\begin{eqnarray}\label{first_ansatz}
   I_{ct} & = & \int_{\dM} \nts dx \, \sqrt{\tg} \, F(\phi(\rw),\tg_{tt}(\rw))  
\end{eqnarray}
The function $F(\phi(\rw),\tg_{tt}(\rw))$ in the integrand is a scalar that depends only on quantities intrinsic to the boundary; i.e. not on normal derivatives of any of the fields. Since the function $F$ is a scalar, and there are no non-trivial curvature invariants intrinsic to the one-dimensional boundary, we conclude that it cannot have any dependence on the induced metric $\tg_{tt}$. Furthermore, derivatives of $\phi(\rw)$ along the boundary vanish, so  $F(\phi(\rw))$ can only be a function of the dilaton at the boundary. With this restriction, the Hamilton-Jacobi equation actually reduces to a \emph{linear} differential equation for the quantity $F(\phi(\rw))^2$. Solving this equation would give a one-parameter family of solutions $F(\phi(\rw))^2$.
However, we have not yet exhausted all of the symmetries of the action. Namely, the action \eqref{bosonic_action} is invariant under the duality transformation \cite{Buscher:1987sk,Buscher:1987qj}:
\begin{eqnarray}\label{duality}
	g_{tt} & \rightarrow & \frac{1}{g_{tt}} \\ \nonumber
	\phi   & \rightarrow & \phi -\frac{1}{2}\, \log(|g_{tt}|)
\end{eqnarray} 
We want to make sure that the counterterm contributions to the action respect this symmetry as well. The conclusion is that the integrand of \eqref{first_ansatz} should be invariant under \eqref{duality}. The only ansatz consistent with boundary diffeomorphism invariance and the T-duality transformation \eqref{duality} is:
\begin{eqnarray}\label{bosonic_ansatz}
	I_{ct} & = & b \, \int_{\dM} \nts dx \, \sqrt{\tg} \, e^{-2 \phi}  
\end{eqnarray}
where $b$ is a constant which is determined by solving the Hamilton-Jacobi equation.
Using the definitions \eqref{bosonic_momenta} and the ansatz \eqref{bosonic_ansatz}, the momenta are:
\begin{eqnarray}
   \pi_{\phi} & = & -2 \,b\,e^{-2\phi} \\
   \pi^{tt}   & = & -\frac{1}{2\,l(r)}\,b\,e^{-2\phi}
\end{eqnarray}
Solving the Hamilton-Jacobi equation \eqref{Hamilton_Jacobi_Equation} determines\footnote{Since the equation is quadratic, it only determines $b$ up to a sign. Only one choice for $b$ removes divergences from the action. Alternately, one may motivate that choice of $b$ by noting that, in the variation of the renormalized action $\Gamma$, momentum flow across the boundary is zero due to counterterm contributions, so that the action is properly extremized for the finite region inside the boundary.} the constant $b$, which gives 
the counterterm action
\begin{eqnarray}\label{bosonic_ct_action}
	I_{ct} & = & -2\,\sqrt{c} \, \int_{\dM} \nts \nts dx \, \sqrt{\tg} \, e^{-2 \phi}  
\end{eqnarray}
The renormalized action is therefore given by:
\begin{eqnarray}\label{Witten_Renormalized_Action}
 \Gamma & = & -\beta \, \sqrt{c} \, e^{-2 \phi(\rw)} \, \left( 1 -2 \sqrt{l(\rw)} + l(\rw) \right)
\end{eqnarray}

\subsection{Does It Work?}\label{subsec:Does_It_Work}
To verify that \eqref{Witten_Renormalized_Action} is, in fact, the correct action to use in studying the thermodynamics of Witten's black hole, we now show that it reproduces the results of \cite{Gibbons:1992rh}. Using the Tolman relationship \eqref{Tolman_factor} and the thermodynamic variables discussed in section \ref{sec:Ensembles}, the renormalized action corresponds to a Helmholtz free energy:
\begin{eqnarray}
 F_{W}(T_W,D_W) & = & -\sqrt{c} \, D_W \, \left( 2 - \frac{T}{T_W} - \frac{T_W}{T}\right)
\end{eqnarray}
Note that the counterterm contribution gives a term proportional to the dilaton charge, but does not depend on either $T$ or $T_W$. We can now use the thermodynamic relations \eqref{canonical_defns} to obtain the entropy and chemical potential. The central charge is always taken to be constant, and therefore derivatives of $T$ will vanish by \eqref{bosonic_T_c_reln}. After a little algebra, one obtains the entropy:
\begin{eqnarray}
   S & = & - \left. \frac{\partial F_W}{\partial T_W} \, \right\vert_{D_W} \\ \nonumber
       & = & 4\pi\,e^{-2\phi(\rh)}
\end{eqnarray}
This agrees with the standard result \cite{Gibbons:1992rh}. The exponential is just the dilaton charge evaluated at the horizon. Thus:
\begin{eqnarray}\label{Witten_BH_Entropy_DH}
   S & = & 4 \pi \,D_H
\end{eqnarray}
It is important to note that, unlike the dilaton charge $D_W$, the entropy is independent of where the wall is, because the only contribution comes from the black hole. 
The chemical potential is:
\begin{eqnarray}
   \psi_W & = &  \left. \frac{\partial F_W}{\partial D_W} \, \right\vert_{T_W} \\ \nonumber
          & = & \sqrt{c} \, \left(2 - \frac{T}{T_W} - \frac{T_W}{T} \right)
\end{eqnarray}
This result is in agreement with the observation that, for a system with a single conserved charge $D_W$, the Helmholtz free energy should be given by:
\begin{eqnarray}
   F_W & = & \psi_W \, D_W  
\end{eqnarray}
Finally, using the relationship between the energy and the Helmholtz free energy gives the mass of the black hole spacetime:
\begin{eqnarray}\label{Witten_BH_Mass}
  M_W & = & 2 \sqrt{c} \, D_W \, \left( 1-\frac{T}{T_W} \right)
\end{eqnarray}
Because this system is particularly simple, it is straightforward\footnote{In section \ref{subsec:First_Law} we will demonstrate the first law for the 0A black hole with constant RR flux. The $q \rightarrow 0$ limit reproduces the first law for Witten's black hole.} to verify the first law of black hole thermodynamics:
\begin{eqnarray}
  d M_W & = & T_W \, dS + \psi_W \, dD_W
\end{eqnarray}
In the limit in which the wall is taken to infinity, the mass of the spacetime~\footnote{Of course, as with all
theories of gravity, there are natural ambiguities in the definition of a mass. We take the point of view that these
ambiguities are fixed by consistency with the first law of thermodynamics and the quantum statistical relation. For
thorough discussions on the notion of mass in 2D dilaton gravity, see \cite{Grumiller:2002nm,Grumiller:2004wi} and
references therein.}
asymptotes to the black hole mass:
\begin{eqnarray}
  M_{BH} & = & \sqrt{c}\,D_H
\end{eqnarray}
All of these results are in complete agreement with those of \cite{Gibbons:1992rh}. Furthermore, the last result agrees with Mann's definition of the mass \cite{Mann:1992yv}, which generalizes earlier work of Frolov \cite{Frolov:1992xx}.

The contribution of the counterterm action is now clear. Because it modified the free energy by a term that did not depend on the temperature, it did not affect the calculation of the entropy. Rather, the counterterm action served to renormalize what would have been a finite chemical potential for the exponentially growing dilaton charge. As a result, both the renormalized action and the mass of the black hole are finite.

\subsection{The 0A Black Hole}

The results of the previous section agree with the usual treatment of the Witten black hole using background subtraction. This inspires confidence in the counterterms determined by the Hamilton-Jacobi method. We now apply this technique to the case of the $0A$ black hole with constant RR flux.  

Our ansatz for the $0A$ solution is identical to the ansatz of the previous section. The only difference is in the justifications for this particular form. As before, we propose a general ansatz, and refine it via considerations of symmetries. We want a counterterm action that is invariant under diffeomorphisms of the boundary coordinate. Since there are no non-trivial intrinsic curvature invariants which we can construct from the induced metric, we expect a counterterm action of the form:
\begin{eqnarray}\label{first_0A_ansatz}
   I_{ct} & = & \int_{\dM} \nts dx \, \sqrt{\tg} \, F(\phi(\rw),A_t(\rw))  
\end{eqnarray}
The only local gauge-invariant quantities which can be constructed from the gauge field $A_t$ 
are the field strength $F_{rt}$, or invariants built from gauge covariant derivatives of charged fields. 
The field strength involves a normal derivative of $A_t$ and is, therefore, not intrinsic to the boundary. Including it would modify the bulk equations of motion, so any dependence on $A_t$ via the field strength is ruled out. Since the dilaton is neutral we cannot introduce $A_t$ through gauge covariant derivatives, either. We once again find that the counterterm action must be of the form:
 \begin{eqnarray}\label{second_0A_ansatz}
   I_{ct} & = & \int_{\dM} \nts dx \, \sqrt{\tg} \, F(\phi(\rw))  
\end{eqnarray}

In the previous section we used the duality transformations \eqref{duality} to pin down the form of the function $F(\phi(\rw))$. A generic solution of the $0A$ theory is not invariant under this transformation, but there is a T-duality relating solutions of the $0A$ and $0B$ theories which leaves the value of the on-shell action unchanged. Specifically, the $0A$ solution with constant RR flux is T-dual to a solution of type $0B$ with a null one-form field strength \cite{Gukov:2003yp}. Furthermore, the transformation of the NS-NS fields $\phi$ and $\tg_{tt}$ are exactly the same as the duality transformation we've already considered. Since the counterterm action only depends on the NS-NS fields $\phi$ and $\tg_{tt}$, it should be invariant under the T-duality. This leads to the same form of the counterterm action that we found in the previous section:
\begin{eqnarray}\label{0A_ct_action}
   I_{ct} & = & -2\,\sqrt{c} \, \int_{\dM} \nts \nts dx \, \sqrt{\tg} \, e^{-2 \phi}  
\end{eqnarray}
Adding this contribution to the on-shell action \eqref{OA_reg_action} gives the renormalized on-shell action:
\begin{eqnarray}
  \Gamma & = &  -\beta \, \sqrt{c} \, e^{-2 \phi(\rw)} \, \left( 1 -2 \sqrt{l(\rw)} + l(\rw) \right) + 
  		\beta \, \frac{q^2}{2\pi\alpha'\sqrt{c}}-\beta \frac{q^2}{2\pi\alpha'}\,(\rw-\rh)
\end{eqnarray}
The corresponding thermodynamic potential in the mixed ensemble is:
\begin{eqnarray}\label{0A_H}
  Y_W(T_W,D,\Phi_W) & = & \sqrt{c} \, \frac{T_W}{T} \, D_H - 4\pi T \, D_H 
  	+ \sqrt{c}\,D\,\left(2-\frac{T_W}{T}-\frac{T}{T_W} \right) - q\,\Phi_W
\end{eqnarray}
where it is understood that $D_H$ and $T$ depend implicitly on the thermodynamic variables $T_W$,$D_W$, and $\Phi_W$.
In obtaining this expression we used the relationship \eqref{0A_temp_reln}. Because this system possesses more than one conserved charge its thermodynamics is more complicated than the previous example's. We devote the next section to understanding the thermodynamics of the 0A black hole spacetime.

It is important to point out that the renormalized action for the 0A black hole contains a term which diverges in the $\rw \rightarrow \infty$ limit. This is in contrast to higher dimensional examples, in which $\Gamma$ tends to be finite, with the exception of possible log divergences related to conformal anomalies. Specifically, because the electrostatic potential is linear in two dimensions both the renormalized action and the mass of the spacetime contain a term linear in $(\rw-\rh)$, the separation between the wall and the horizon. A divergence of this sort as $\rw \rightarrow \infty$ is not unphysical, it is merely a consequence of two-dimensional electrostatics. Misinterpreting the nature of this divergence and attempting to remove it from the renormalized action $\Gamma$ leads to thermodynamic quantities which are \emph{not} consistent with the first law.

\section{The Thermodynamics of the 0A Black Hole}

In this section we use the thermodynamic potential \eqref{0A_H} to determine the entropy, dilaton chemical potential, electric charge, and mass for the 0A black hole. As a check of our work we demonstrate that the first law of black hole thermodynamics is satisfied. We point out how this result is in agreement with Mann's result for the mass of a 2d black hole, and explain how it disagrees with several other results in the literature.

\subsection{The Entropy}
The entropy is given by:
\begin{eqnarray}
   S_W & = & - \left. \frac{\partial Y_W}{\partial T_W} \, \right\vert_{D_W,\Phi_W}
\end{eqnarray}
To evaluate this expression we need to determine the conditions associated with holding $D_W$ and $\Phi_W$ constant:
\begin{equation}
    \frac{\partial D_W}{\partial T_W} \, \, = \, \, 0 \,\,~~~~~~~ \frac{\partial \Phi_W}{\partial T_W} \, \, = \, \, 0
\end{equation}
These two equations lead to the following conditions:
\begin{eqnarray}
\frac{\partial}{\partial T_W} \, (\rw-\rh) & = & -\frac{1}{\sqrt{c}} \, D_H^{-1}\,\frac{\partial D_H}{\partial T_W}
\end{eqnarray}
\begin{eqnarray}
\Phi_W \, \frac{\partial q}{\partial T_W} & = & \frac{q^2}{\sqrt{c}}\,D_H^{-1}\,\frac{T_W}{T}\,\frac{\partial D_H}{\partial T_W}
	+ q\,\Phi_W \, \left( \frac{1}{T}\,\frac{\partial T}{\partial T_W} - \frac{1}{T_W} \right) 
\end{eqnarray}
Using these expressions, along with the identity \footnote{This is obtained from the Tolman relation \eqref{Tolman_factor} by rewriting $g_{tt}$ in terms of $D_W$,$D_H$,$q$, and $\Phi$.}:
\begin{eqnarray}
D_H & = & D_W \,\left(1-\frac{T^2}{T_W^2} \right) - \frac{1}{\sqrt{c}}\,q\,\Phi
\end{eqnarray}
the entropy is determined to be:
\begin{eqnarray}\label{0A_entropy}
   S & = & 4 \pi D_H
\end{eqnarray}
The entropy is in agreement with the results found in \cite{Gukov:2003yp,Davis:2004xb, Danielsson:2004xf}. It is independent of the position of the wall, as expected, and has the same functional form as the entropy of Witten's black hole. This suggests a reassuring analogy with higher dimensional black holes, where the entropy is always found to be one quarter of the horizon area, in Planck units, regardless of whether or not the black hole carries an electric charge.

\subsection{The Electic Charge}
The total electric charge of the solution is obtained from:
\begin{eqnarray}
   Q & = & - \left. \frac{\partial Y_W}{\partial \Phi_W} \, \right\vert_{D_W,T_W}
\end{eqnarray}
As with the evaluation of the entropy, it is necessary to find the conditions associated with holding $D_W$ and $T_W$ fixed:
\begin{eqnarray}
\frac{\partial}{\partial \Phi_W} \, (\rw-\rh) & = & -\frac{1}{\sqrt{c}} \, D_H^{-1}\,\frac{\partial D_H}{\partial \Phi_W}
\end{eqnarray}
\begin{eqnarray}
\Phi_W \, \frac{\partial q}{\partial \Phi_W} & = & q + \frac{T_W}{T} \, \frac{q^2}{\sqrt{c}}\,D_H^{-1}\,
	\frac{\partial D_H}{\partial \Phi_W} + \frac{1}{T} \, q \, \Phi_W \, \frac{\partial T}{\partial \Phi_W}
\end{eqnarray}
To obtain the second relationship we have also used \eqref{Phi_W_definition}, the definition of $\Phi_W$. Using these expressions, a small amount of algebra leads to the result:
\begin{eqnarray}\label{0A_electric_charge}
   Q & = & 2 q
\end{eqnarray}
At this point it is important to remember that we have been working with a non-canonically normalized field strength which actually represents two canonical field strength terms in the action, both with flux $q$. Thus, the total conserved electric charge of $2q$ is merely a conserved charge $q$ for each gauge field. It is important to remember the origin of this factor of 2 when identifying the contribution of the electric fields to the total mass of the black hole spacetime.

\subsection{The Dilaton Charge Chemical Potential}
The chemical potential associated with the dilaton charge is given by:
\begin{eqnarray}
   \psi_W & = &  \left. \frac{\partial Y_W}{\partial D_W} \, \right\vert_{T_W,\Phi_W}
\end{eqnarray}
The two identities associated with holding $T_W$ and $\Phi_W$ fixed are:
\begin{eqnarray}
 \frac{\partial}{\partial D_W}\,\left( \rw-\rh\right) & = & \frac{1}{\sqrt{c}}\,D_W^{\,-1} - \frac{1}{\sqrt{c}}\,D_H^{-1} \, 
 	\frac{\partial D_H}{\partial D_W} 
\end{eqnarray}
\begin{eqnarray}
  \Phi_W \, \frac{\partial q}{\partial D_W} & = & \frac{1}{T} \, \frac{\partial T}{\partial D_W}\,q\,\Phi_W - \frac{T_W}{T} \,   
  \frac{q^2}{\sqrt{c}} \, D_W^{\,-1} + \frac{T_W}{T}\,\frac{q^2}{\sqrt{c}}\,D_H^{-1}\,\frac{\partial D_H}{\partial D_W}
\end{eqnarray}
Using these expressions, the chemical potential is:
\begin{eqnarray}\label{0A_chem_potential}
  \psi_W & = & \sqrt{c}\,\left( 2-\frac{T}{T_W}-\frac{T_W}{T}\right) + \frac{T_W}{T}\,\frac{q^2}{\sqrt{c}}\,D_W^{\,-1} 
\end{eqnarray}

\subsection{The Mass of the 0A Black Hole}
Now that we have obtained the entropy, electric charge, and dilaton potential, it is straightforward to obtain the mass (energy) of the 0A black hole spacetime. The partition function was calculated in a mixed ensemble. From the point of view of the dilaton charge it is the canonical ensemble, while from the point of view of the electric charge it is the grand canonical ensemble. Therefore, the Helmholtz free energy is related to the log of the partition function, $Y_W$, by an inverse Legendre transform:
\begin{eqnarray}
  F_W(T_W,D_W,Q) & = & Y_W(T_W,D_W,\Phi_W) + Q\,\Phi_W
\end{eqnarray}
Using the standard relationship between the energy and the Helmholtz free energy, we obtain:
\begin{eqnarray}
  M_W & = & Y_W + T_W\,S + Q\, \Phi_W
\end{eqnarray}
Using the results of the previous sections, the energy is given by:
\begin{eqnarray}\label{0A_Mass}
  M_W & = & 2\sqrt{c}\,D_W\,\left( 1-\frac{T}{T_W}\right)
\end{eqnarray}
The functional form of the energy is identical to that of Witten's black hole \eqref{Witten_BH_Mass}. However, the asymptotic value of the total energy of the spacetime is quite different, due to the presence of the linear electric potential associated with the RR charge. Expanding \eqref{0A_Mass} for $\rw\gg\rh$ we find:
\begin{eqnarray}\label{asymptotic_0A_Mass}
  M_W & \approx & \sqrt{c}\,D_H + q \, \Phi + \OO(e^{-\sqrt{c}\,\rw})
\end{eqnarray}
As we move the wall further away from the horizon, the total energy decomposes into two distinct contributions, each with a clear physical interpretation. The first term is the mass of the black hole itself:
\begin{eqnarray}\label{0A_BH_Mass}
   M_{BH} & = & \sqrt{c} \, D_H
\end{eqnarray}
and the second term is merely the acculumated electrostatic energy between the wall and the horizon:
\begin{eqnarray}
   M_{\Phi} & = & \int_{\rh}^{\rw} \nts \nts dr \, (F_{rt})^2 \\
   	    & = & q\,\Phi
\end{eqnarray}
The integrand is the electrostatic energy density associated with our non-canonically normalized field strength.

There are several reasons to believe that this result is correct. The first is that the mass of the black hole agrees with the definition proposed by Mann \cite{Mann:1992yv}. The second, which we will turn to momentarily, is that our result \eqref{0A_Mass} is consistent with the first law of black hole thermodynamics. But before we proceed, it is worth pointing out that our result for the mass \emph{disagrees} with a number of results in the literature \cite{Berkovits:2001tg,Gukov:2003yp,Davis:2004xb}. For example, a value that is often quoted for the black hole mass is:
\begin{eqnarray}\label{alternate_mass}
  M' & = & 4\,\frac{m}{\sqrt{c}} 
\end{eqnarray}
where the parameter $m$ appeared in \eqref{alternate_0A_metric}. Using \eqref{extremality_parameter} we can rewrite this in our conventions as:
\begin{eqnarray}
  M' & = & \sqrt{c}\,D_H - \frac{q^2}{2\pi\alpha'}\,\rh
\end{eqnarray}
This is what one would obtain from \eqref{asymptotic_0A_Mass} by discarding the `divergent' term linear in $\rw$. As was pointed out earlier, this term is due to the linear electrostatic potential and should not be discarded. The resulting mass \eqref{alternate_mass} is not consistent with the first law of black hole thermodynamics.

\subsection{The First Law}\label{subsec:First_Law}

As a verification of our result \eqref{0A_Mass} for the mass of the 0A black hole spacetime, we show that it satisfies the first law of black hole thermodynamics:
\begin{eqnarray}
  dM_W & = & \psi_W \, dD_W + T_W \, dS + \Phi_W \,dQ
\end{eqnarray}
This sort of calculation is usually not shown explicitly, as it often reduces to an excercise in algebra and differentials. However, since our expression \eqref{0A_Mass} for the mass disagrees with several other results, we feel it is necessary to demonstrate the first law in detail. Also, note that setting $q$ to zero throughout this section gives a proof of the first law for Witten's black hole, as mentioned in section \ref{subsec:Does_It_Work}.

We begin by taking the differential of the expression \eqref{0A_Mass}
\begin{eqnarray}
  dM_W & = & 2\,\sqrt{c}\,\left( 1-\frac{T}{T_W} \right)\,dD_W + 2\sqrt{c}\,D_W \,d\left( 1-\frac{T}{T_W}\right)
\end{eqnarray}
By adding and subtracting factors of $T/T_W$ and $T_W/T$ this can be re-written in the form:
\begin{eqnarray} \nonumber
  dM_W & = & \sqrt{c}\,\left( 2-\frac{T}{T_W} -\frac{T_W}{T}\right)\,dD_W + \sqrt{c}\,\frac{T_W}{T}\,
  \left(1-\frac{T^2}{T_W^{2}} \right)\,dD_W - \sqrt{c}\,D_W\,\frac{T_W}{T}\,d\left( \frac{T^2}{T_W^2}\right) \\ \label{step_3}
       & = & \sqrt{c}\,\left( 2-\frac{T}{T_W} -\frac{T_W}{T}\right)\,dD_W + \sqrt{c}\,\frac{T_W}{T}\,
       d\left( D_W\left(1-\frac{T^2}{T_W^2}\right)\right)
\end{eqnarray}
Next we need the result \eqref{0A_chem_potential} for the dilaton chemical potential $\psi_W$, as well as the identity:
\begin{eqnarray}
   D_W\,\left( 1-\frac{T^2}{T_W^2}\right) & = & D_H + \frac{1}{\sqrt{c}}\,q\,\Phi
\end{eqnarray}
Using these two expressions in \eqref{step_3} allows us to write $dM_W$ as:
\begin{eqnarray}
  dM_W & = & \psi_W \, dD_W + \sqrt{c}\,\frac{T_W}{T}\,d D_H + \frac{T_W}{T}\, d\left( q\,\Phi\right)
  	-\frac{T_W}{T}\,\frac{q^2}{\sqrt{c}}\,d\log{D_W}
\end{eqnarray}
Using the relationship \eqref{0A_temp_reln} between $T$,$c$,$q$, and $D_H$ we can expand the second term on the right hand side, which leads to:
\begin{eqnarray}
  dM_W & = & \psi_W\,dD_W + T_W d\left( 4\pi\,D_H\right) + \frac{T_W}{T}\,q\,d\Phi + 
  	\Phi_W\,dq - \frac{T_W}{T}\,\frac{q^2}{\sqrt{c}}\,d\log{\left(\frac{D_W}{D_H}\right)}
\end{eqnarray}
Referring to the result \eqref{0A_entropy} for the entropy, the second term is just $T_W\,dS$. Since the total electric charge is $Q=2q$, we can rewrite the differential as:
\begin{eqnarray}\label{next_to_last_step}
  dM_W & = & \psi_W \,dD_W + T_W \, dS + \Phi_W \, dQ + \frac{T_W}{T}\,q\,d\Phi-\Phi_W\,dq -
   \frac{T_W}{T}\,\frac{q^2}{\sqrt{c}}\,d\log{\left(\frac{D_W}{D_H}\right)}
\end{eqnarray}
Using the expression \eqref{dilaton_charge} for $D_W$ the last term on the right hand side is:
\begin{eqnarray}
 - \frac{T_W}{T}\,\frac{q^2}{\sqrt{c}}\, d\log{\left(\frac{D_W}{D_H}\right)} & = & 
 	- \frac{T_W}{T}\,q^2\,d\left(\rw-\rh\right) \\ \nonumber
	& = & -\frac{T_W}{T}\,q\,\left( d \Phi -\left(\rw-\rh\right)\,dq \right) \\ \nonumber
	& = & -\frac{T_W}{T}\,q\,d\Phi + \Phi_W \, dq
\end{eqnarray}
Thus, the last three terms in \eqref{next_to_last_step} cancel, leaving:
\begin{eqnarray}
  dM_W & = & \psi_W \, dD_W + T_W\,dS + \Phi_W \, dQ
\end{eqnarray}
This confirms that our results for the 0A black hole spacetime satisfy the first law of black hole thermodynamics. We should also point out that the first law, as shown here, is true regardless of the location of the regulating surface.


\section{Discussion}

In this paper we have analyzed the thermodynamics of black hole solutions with constant RR flux in two-dimensional type 0A string theory. After clarifying the role of boundary conditions in determining the thermodynamic ensemble we calculated the renormalized, on-shell actions that determine the partition function. The approach we used to remove infra-red divergences from the partition function is based on similar techniques used to address infra-red divergences in gravitational actions in higher dimensions \cite{Balasubramanian:1999re,Emparan:1999pm}. Unlike background subtraction, our approach does not require the identification of a reference spacetime. We motivated this technique by applying it to the black hole studied by Witten in \cite{Witten:1991yr}. The results we obtain for the 0A black hole provide a consistent picture of the thermodynamics of these solutions. In particular, our result \eqref{asymptotic_0A_Mass} for the mass of the black hole spacetime resolves a number of contradictory claims in the literature. We are confident in our results, as they are the only ones (of which we are aware) that satisfy the first law of black hole thermodynamics. 

From the semiclassical point of view our results do not support the conclusions of proposed matrix model duals of the $0A$ black hole. However, this does not necessarily invalidate the matrix model conjectures. A necessary caveat for our analysis is that the semiclassical description may not sufficiently capture the physics of two-dimensional black holes. Because curvatures near the horizon are of order $\OO(1/\alpha')$ the semiclassical analysis will generically break down there, as stringy corrections become important.

For this reason, two-dimensional black holes may be too stringy to have a valid geometric limit at all. In that case, one needs separate techniques on the string/gravity side in order to verify matrix model predictions. For example, Witten's black hole can be described using a gauged WZW coset model. If we set the string coupling in the region exterior to the horizon to be small, $g_H = \exp \left( \phi_H \right) \ll 1$, then quantum corrections can be controlled and the genus zero analysis is sufficient. String corrections, however, may be important. To see this, note that the action \eqref{bosonic_action} corresponds to the $k \rightarrow \infty$ limit of the worldsheet theory. The expansion in $1/k$ is equivalent to the $\alpha'$ expansion.  But the two dimensional string is critical for the relatively small value  $k=9/4$, which means that the spacetime effective action should receive significant $\alpha'$ corrections. 
A solution exists which incorporates higher order corrections \cite{Dijkgraaf:1991ba}, but the corresponding effective action, which is necessary to calculate the thermodynamic potential, is not known. It would be interesting to see whether or not the method used here could be applied to the considerations of \cite{Kazakov:2001pj}, perhaps extending our analysis to finite level $k$.
It is not clear what aspects of the critical string black hole are captured by the large level analysis. It is interesting to note that there are $D=10$ critical string backgrounds which contain Witten's black hole as a two-dimensional subspace, for which a sensible large $k$ limit exists. Our analysis may be completely valid in such cases. But, for now, a worldsheet theory describing the 0A black hole with constant RR flux is not known to exist.

The results derived in this paper support the idea that two dimensional black holes, like their higher dimensional analogues, seem to exhibit certain universal features. The most obvious example is the entropy. In both the charged and uncharged cases we find that the entropy takes the form:
\begin{equation}\label{Disc_BH_Entropy}
   S ~~ = ~~ 4 \pi D_H ~~ = ~~ \frac{4\pi}{g_H^{\,2}}
\end{equation}
This is analagous to higher dimensional black holes, where the entropy is always given by a quarter of the horizon area, measured in Planck units:
\begin{eqnarray}
   S & = & \frac{A_H}{4 \, G}
\end{eqnarray}
In fact, \eqref{Disc_BH_Entropy} is equivalent to this formula. Our calculations were all performed in units where $2\kappa^2 = 16 \pi G = 1$. Restoring these units, we have:
\begin{eqnarray}
  S & = & \frac{D_H}{4 G}
\end{eqnarray}
The area of a `sphere' with coordinate radius $r$ in $n$ dimensions is given by:
\begin{eqnarray}
  A(r) & = & \frac{2\,\pi^{n/2}}{\Gamma\left(\frac{n}{2}\right)}\,r^{n-1}  
\end{eqnarray}
so the area of the horizon in one spatial dimension is simply $A=2$, independent of $\rh$. An observer outside the horizon has access to only one of the two points composing the 1-sphere, so the area that should appear in the entropy-area relation is $A_H=1$. Thus, \eqref{Disc_BH_Entropy} is in fact:
\begin{eqnarray}
  S & = & \frac{A_H}{4 G_H}
\end{eqnarray}
where $G_H$ is the effective two dimensional Newton's constant at the horizon:
\begin{eqnarray}
  G_H & = & G \, e^{2 \phi_H}
\end{eqnarray}
Because this quantity is dimensionless in two dimensions, it is more appropriate to regard it as setting an effective Planck resolution which defines the notion of `per unit area'. Notice that as we decrease the string coupling the black hole becomes both more entropic and more massive, which is in agreement with our intuition that making the system very massive should make quantum corrections less important.  

It is also interesting that the total mass of the spacetime out to the wall takes the same form in both the charged and uncharged cases:
\begin{eqnarray}
  M_W & = & 2\sqrt{c}\,D_W\,\left( 1-\frac{T}{T_W}\right)
\end{eqnarray}
In the case of Witten's black hole this quantity asymptotes to the black hole mass $\sqrt{c}\,D_H$ as $\rw \rightarrow \infty$. In the 0A case its asymptotic form comprises two distinct contributions, the black hole mass and an electrostatic energy due to the electric field.

Finally, the counterterm calculated in section \ref{sec:Divergences} captures a universal infra-red divergence in linear dilaton gravitational backgrounds. The effect of the counterterm action can be thought of as subtracting a divergent contribution from the thermodynamic potential (and mass of the spacetime) related to the dilaton charge. This should be expected; the dilaton is a long range scalar field that violates the weak equivalence principle. It was pointed out in \cite{Gibbons:1992rh} that the energy associated with the term we subtract does not gravitate; it is an irreducible energy due to the dilaton. It is remarkable that such a simple procedure isolates this divergence. Once we made an ansatz of local terms on the regulating boundary, specifying the symmetries of the counterterm action was sufficient to completely specify its functional form.

Our approach can be viewed as validating the use of background subtraction by reproducing its results without having to appeal to a reference spacetime. In the same vein, we provide an alternate justification for the subtractions considered in \cite{McGuigan:1991qp,Nappi:1992as}. One of the conclusions of this paper is that the Hamilton-Jacobi method of determining boundary counterterms seems to be applicable to a wide range of interesting gravitational backgrounds.

\section*{Acknowledgements}
The authors would like to thank Ben Burrington, Jason Kumar, Finn Larsen, Jim Liu, Leo Pando-Zayas, Malcolm Perry, and especially Diana Vaman for helpful conversations. In addition, we would like to thank Malcolm Perry for drawing our attention to several references and for being kind enough to read an early version of this paper. This research was supported in part by the DoE through Grant No.
DE-FG02-95ER40899 (Michigan).

\pagebreak

\end{document}